# Rotation of the dislocation grid in multilayer FeSe films and visualization of electronic nematic domains via orbital-selective tunneling

Zheng Ren[1*], Hong Li[1], He Zhao[1], Shrinkhala Sharma[1] and Ilija Zeljkovic[1†]

[1]Department of Physics, Boston College, 140 Commonwealth Ave, Chestnut Hill, MA 02467, USA

*zr10@rice.edu

†ilija.zeljkovic@bc.edu

**Abstract**

Understanding the interplay of structural and electronic symmetry breaking in Fe-based high temperature superconductors remains of high interest. In this work we grow strain-patterned multilayer FeSe thin films in a range of thicknesses using molecular beam epitaxy. We study the formation of electronic nematic domains and spatially-varying strain using scanning tunneling microscopy and spectroscopy. We directly visualize the formation of edge dislocations that give rise to a two-dimensional edge dislocation network in the films. Interestingly, we observe a 45 degree in-plane rotation of the dislocation network as a function of film thickness, yielding antisymmetric strain along different directions. This results in different coupling ratios between electronic nematic domains and antisymmetric strain. Lastly, we are able to distinguish between different orthogonal nematic domains by revealing a small energy-dependent difference in differential conductance maps between the two regions. This could be explained by orbital-selective tip-sample tunneling. Our observations bring new insights into the dislocation network formation in epitaxial thin films and provide another nanoscale tool to explore electronic nematicity in Fe-based superconductors.

**Introduction**

Spontaneous breaking of the rotational symmetry of the electronic structure is widely observed in many unconventional electron systems [1–24]. In Fe-based superconductors (Fe-SCs), electronic nematic ordering that breaks the rotational symmetry of the lattice is nearly always accompanied by tetragonal-to-orthorhombic structural phase transition [12–14]. It manifests itself by various electronic signatures exhibiting a two-fold rotational symmetry ($C_2$), such as the splitting of the bands with $d_{xz}$ and $d_{yz}$ orbital characters observed by angle-resolved photoemission spectroscopy (ARPES) [3,4,15–17], and the resistivity anisotropy along the two inequivalent lattice directions [18,19]. In scanning tunneling microscopy/spectroscopy (STM/S) measurements, $C_2$-symmetric electron scattering has been used as a smoking gun signature of electronic nematicity [20–24].

While the measurements of the Fe-SC bulk single crystals set the foundation for understanding many aspects of the electronic nematic phase, synthesis of FeSC thin films provides a playground for tuning the electronic nematic phase, sometimes to a degree that cannot be achieved in bulk crystals. For example, in ultrathin FeSe/SrTiO$_3$ films, the Pomeranchuk nematic order [25] and a smectic phase [26] have been reported to emerge. In our previous report [27], we have also shown that epitaxially-grown FeSe thin films can exhibit a two-dimensional modulation network, resulting in a spatially-varying antisymmetric strain field with a nanoscale wavelength. Such strain field has a direct impact on the formation of electronic nematic domains due to the nemato-elastic coupling. Here, we synthesize FeSe multilayer films with different thickness using molecular beam epitaxy (MBE) and study the modulation network formation using low temperature STM/S. We observe a remarkable evolution of the dislocation network as a function of thickness. Specifically, the characteristic direction of the network rotates away from the Fe-Fe lattice direction for thinner films, towards the Se-Se lattice direction for thicker films. The modulation spacing additionally gradually changes with thickness. Such structural evolution has direct impact on the distribution of the electronic nematic domains. Interestingly, we also use orbital-selective STM tips to identify the nematic domains. In contrast to other STM signatures of nematicity where a large field of view or specific impurities are required to generate the $C_2$-symmetric electron scattering signal [20–24,28], our method works at the nanoscale and can in principle be extended to other Fe-based superconductors without the need of impurities or defects.

**Methods**

FeSe films were synthesized on Nb-doped (0.05 wt%) SrTiO$_3$(001) substrates (Shinkosha) in our home MBE system (Fermion Instruments). The substrate pre-treatment, growth condition and sample transfer method are described in our previous report [27]. To synthesize films with different thicknesses, we simply vary the time of the growth. The film thickness was determined by measuring the height of the film surface from bare substrates observed in the vicinity.

STM data was acquired using a customized Unisoku USM1300 STM at about 4.5 K. Spectroscopic data was acquired using a standard lock-in technique with 915 Hz frequency and bias excitation described in figure captions. STM tips used were home-made tungsten tips chemically-etched and annealed in UHV to bright orange color prior to STM measurements.

**Results**

We synthesize the FeSe thin films on Nb-doped (0.05 wt%) SrTiO$_3$(001) substrates using molecular beam epitaxy (MBE) (Methods). Tetragonal FeSe has an in-plane lattice constant $a$ = ~3.77 Å, which results in the lattice mismatch of 3.5% with the SrTiO$_3$(001) substrate. Such a lattice mismatch favors the formation of a two-dimensional modulation network that results in spatially varying strain [27,29–31]. In our previous work [27] we hypothesized that such a modulation network could be a network of misfit dislocations (Fig. 1(b)), but a more direct evidence – the surface defects that are associated with the threading section of the dislocation half-loop – was not observed [32]. Here we study the formation of the 2D modulation network in FeSe films as a function of film thickness. We locate a region where we observe a broken modulation network and a number of surface defects, terminating the modulation lines (Fig. 1(c)-(f)). It is conceivable that the modulation lines are a consequence of the misfit dislocations at the interface, because the surface defects are likely the surface endpoints of the threading dislocations, which should terminate the misfit dislocation lines and form a half-loop. Furthermore, we hypothesize that the misfit dislocations are 90° edge dislocations based on the van der Waals layered structure of the FeSe films, and the fact that no glide steps at the surface are observed anywhere in our STM data. Supporting this picture, in a cousin heterostructure Fe(Te,Se)/SrTiO$_3$(001), scanning transmission electron microscopy (STEM) has revealed that the misfit dislocations are of 90° edge character [33]. However, to conclusively identify the character of the misfit dislocations in our work, STEM measurements will likely be necessary.

We proceed to explore the dislocation networks formed in FeSe thin films as a function of film thickness. We compare the dislocation networks observed in the atomically resolved STM topographs acquired at the surface of a 4 monolayer (ML) film, a 5 ML film and a ~8 ML film (Fig. 2(a)-(c)). Remarkably, there are significant differences between the dislocation networks. First, while the 4 ML network (orange dashed lines in Fig. 2(a)) propagates along the Fe-Fe lattice directions (a,b-axis), the 8 ML network (Fig. 2(c)) is rotated by 45 degrees and is now oriented along the Se-Se lattice directions (x,y-axis). Second, the 4 ML network contains thinner dislocation lines, while the 8 ML dislocation lines appear thicker. Third, as we discuss in the subsequent paragraphs, the average spacing between the dislocation lines gradually decreases with increased thickness.

It is interesting to note that in the 4 ML Fe-Fe network, every other node "protrudes" out from the rest of the network, forming a secondary periodic pattern which is oriented along the Se-Se lattice directions (blue arrows in Fig. 2(a)). This is qualitatively similar to the 5 ML film as well, where we again find that every other modulation crossing is brighter than the rest of the nodes (examples are again denoted by blue arrows). These form a secondary pattern in the Fe-Fe network that stands out even more than in the 4 ML film, while the Fe-Fe dislocation lines (orange dashed lines), though still discernible, are fading.

The approximately 45 degree rotation of the modulation network as a function of thickness and the secondary pattern discussed above are both highly unusual and not immediately expected from theory [32,34]. In particular, elasticity theory predicts that the spacing between the 90° edge dislocation lines $l$ monotonically decreases as the film thickness $h$ increases [35]:

$$l = \frac{b}{f - \frac{b}{8\pi h(1+\nu)}\left(\ln\left(\frac{h}{b}\right)+1\right)}$$

where $b$ is the Burgers vector, $f$ is the lattice mismatch between film and substrate and $\nu$ is the Poisson ratio of the film. While the decrease in $l$ with the increase in $h$ is qualitatively consistent with our work (Fig. 2(a)-(c)), the rotation of the dominant modulation pattern is not captured by the equation above. At the same time, we note that FeSe is structurally complex, with each unit cell consisting of an Fe layer sandwiched between two Se layers. It is conceivable that this secondary pattern formed in thinner films along the Se-Se direction gradually evolves into the 8 ML dislocation network (also oriented along the Se-Se direction) as the film thickness increases. To test this hypothesis, we plot the average spacing between the nearest-neighbor nodes along the Se-Se direction as a function of thickness (Fig. 2(d)). We fit the experimental data with this equation by optimizing $b$, while fixing $f$ at 0.034 (the lattice mismatch between FeSe and SrTiO$_3$) and $\nu$ at 0.18 [36] (Fig. 2(d) solid line). An excellent match between the fitting curve and the experimental data is observed. The optimized Burgers vector is 3.67 Å, remarkably close to the lattice constant of FeSe. This is again consistent with our hypothesis that the misfit dislocations (at least the Se-Se network) are 90° edge dislocations. Therefore, it is conceivable that that the protruding nodes in the 4 ML and 5 ML films form the secondary dislocation network that evolves into the Se-Se dislocation network in thicker films.

From the fitting process described above (Fig. 2(d)), the Burgers vector $b_2$ of the Se-Se dislocation network is found to be exactly one lattice constant of FeSe. This is reasonable because $b_2$ should in principle be the minimum lattice translation vector along the Se-Se lattice direction. Since the spacing $l$ is roughly proportional to the Burgers vector $b$, based on the geometrical relationship of Se-Se and Fe-Fe networks (Fig. 2(a)), the Burgers vector of the Fe-Fe dislocation network $b_1$ should have the magnitude of $b_2/\sqrt{2}$ and be oriented along the Fe-Fe lattice direction, i.e. the lattice constant of the underlying "Fe lattice" or $\frac{1}{2}[110]$ (Fig. 2(e)). However, $\frac{1}{2}[110]$ is not a translation vector of the FeSe lattice, which would indicate that the Fe-Fe dislocations may be imperfect dislocations. This brings up a puzzling question on the origin of the Fe-Fe dislocations. If $\frac{1}{2}[110]$ is the Burgers vector, then the Fe-Fe dislocations might be partial dislocations, although this would bring an additional energy cost associated with the stacking faults induced by partial dislocations [34]. Another factor that plays a role is the interface chemical energy [37]. The balance between chemical interfacial energy and the dislocation energy could contribute to the rotation of the in-plane network, but detailed atomistic simulations for the FeSe/SrTiO$_3$ interface lattice structure, similar to that in Ref. [37], will likely be necessary for a complete understanding of our observations.

Having revealed the surprising rotation of the dislocation network as a function of thickness, we next turn to its impact on the formation of the electronic nematic domains. Electronic nematic domains are identified in STM data by several types of unidirectional electronic features, including unidirectional charge-stripes, C$_2$-symmetric quasiparticle modulation pinned by impurities and C$_2$-symmetric quasiparticle interference pattern [20–24,27,28]. In Fe-SCs, there are two types of orthogonal electronic nematic domains, in which the direction of unidirectional electronic features rotates by 90° between the two [20–24,27,28]. The boundary between any two orthogonal nematic domains results in "worm"-like electronic features seen in for example Fig. 3(c,g) [27,38].

As shown in our previous report on a 4 ML FeSe film, the Fe-Fe dislocation network modulates the formation of the nematic domains, resulting in the decoupling of the structural anisotropy and electronic nematicity in about 25% of the area (Fig. 3(a)-(d)) [27]. Here, we extend this analysis to the Se-Se dislocation network in the 8 ML film (Fig. 3(e)-(h)). In the concomitant dI/dV map, the longer "maze-like" electronic nematic domains are discernible (Fig. 3(g)) [22], in contrast to the patchy domain structure in the 4 ML case (Fig. 3(c)). The noticeable difference in the shape of the nematic domains here from the 4 ML case is due to the Se-Se-oriented dislocation networks in 8 ML, which exerts strain along the Se-Se lattice directions. Since the electronic nematicity couples to the antisymmetric strain along the Fe-Fe lattice directions, larger and more continuous nematic domains are expected in the absence of the Fe-Fe oriented strain. Furthermore, since the pre-existing strain is in the Se-Se directions, the electronic nematicity should dominate the distribution of the Fe-Fe antisymmetric strain. Indeed, a higher coupling of the strain and nematicity is observed in the Fe-Fe antisymmetric strain map (Fig. 3(f),(h)).

By examining the dI/dV(**r**,V) maps of the two orthogonal electronic nematic regions in more detail, we find another surprising distinction. In principle, the two domains are equivalent and as such, they should exhibit identical dI/dV spectra. Remarkably however, we observe a small difference between the average dI/dV spectra in the two orthogonal electronic nematic domains (Fig. 4(b)). This contrast can be more easily distinguished in a dI/dV(**r**,V) map, where it is apparent that the conductance is approximately spatially uniform in each domain, but appears strikingly different when comparing the two (Fig. 4(a)). To explore this further, we acquire a series of dI/dV(**r**,V) maps as a function of bias voltage and plot the difference between the average dI/dV signals in electronic nematic domains A and B (Fig. 4(c)). Interestingly, this dI/dV contrast shows a clear energy dependence. Within the range of the bias voltage explored, between 0 mV and 40 mV, there is a negligible dI/dV difference. Starting from 60 mV, domain A shows a noticeably lower dI/dV signal compared to domain B. The negative offset peaks at around 90 mV. At higher bias voltage of 120-140 mV, the offset reverses, and the average dI/dV signal in domain A is substantially larger than the signal in domain B. We can rule out tip changes giving rise to such a contrast as the data over both domains was acquired using the same STM tip.

The observed contrast in dI/dV conductance between the two orthogonal electronic nematic domains can be easily explained by orbital sensitivity of the STM tip. Previous STM work on FeSe established the use of microscopically different STM tips to enhance sensitivity of the measurement to different electronic bands [28]. In our work, the data can in principle be explained by an STM tip that is more sensitive to one of the two main orbitals that are split in energy: either $d_{xz}$ or $d_{yz}$ orbitals. We note that we have observed the same phenomenon in several FeSe films studied (Fig. S1). We again mention that the small difference in dI/dV signal correlates with the pattern of electronic nematic domains revealed by the unidirectional scattering signature. The energy dependence of the nematic dI/dV contrast roughly supports this picture. In bulk single crystal of FeSe, ARPES measurements have determined a splitting of $d_{xz}$ and $d_{yz}$ orbital degeneracy by ~50 meV [39]. In ~35 ML FeSe thin films, the band splitting is the same order of magnitude and clearly momentum-dependent, with the $d_{yz}$ band pushed even further above the Fermi level [40]. In our ultra-thin films down to a few layer thickness, the energy scale of the dI/dV contrast reversal is roughly consistent with the orbital splitting energy scale. While the details of this physical picture will require further theoretical modeling, our experiments strongly suggest that there exists orbital-selective electron tunneling between the STM tip and the FeSe film, resulting in the difference in measured dI/dV signal between the two nematic domains.

**Conclusion**

Our observation of edge dislocation network that spatially rotates as a function of film thickness provides an intriguing playground to theoretically understand the formation of dislocations in complex systems with dissimilar atomic layers within each unit cell. In addition to using unidirectional quasiparticle interference [20,21,23] and charge-stripe formation [22,24] as the smoking gun signatures of electronic nematicity in STM measurements of FeSe, our work provides another atomic-scale tool to evaluate the degree of electronic nematicity in FeSe. It should also be applicable to other related Fe-based superconductors. To fully understand the energy-dependence of the dI/dV contrast (Fig. 4), future temperature-dependent STM/S measurements of the nematic dI/dV contrast across nematic critical temperature in FeSe and other Fe-SCs would be of highly desirable.

**Acknowledgements**

Z.R. and I.Z. acknowledge the fruitful discussions with Mingda Li, Pratik Dholabhai, Frederick Walker and Charles Ahn. I.Z. acknowledges the support from National Science Foundation grant no. NSF-DMR-2216080 (STM experiments) and DARPA grant no. N66001-17-1-4051 (MBE synthesis).

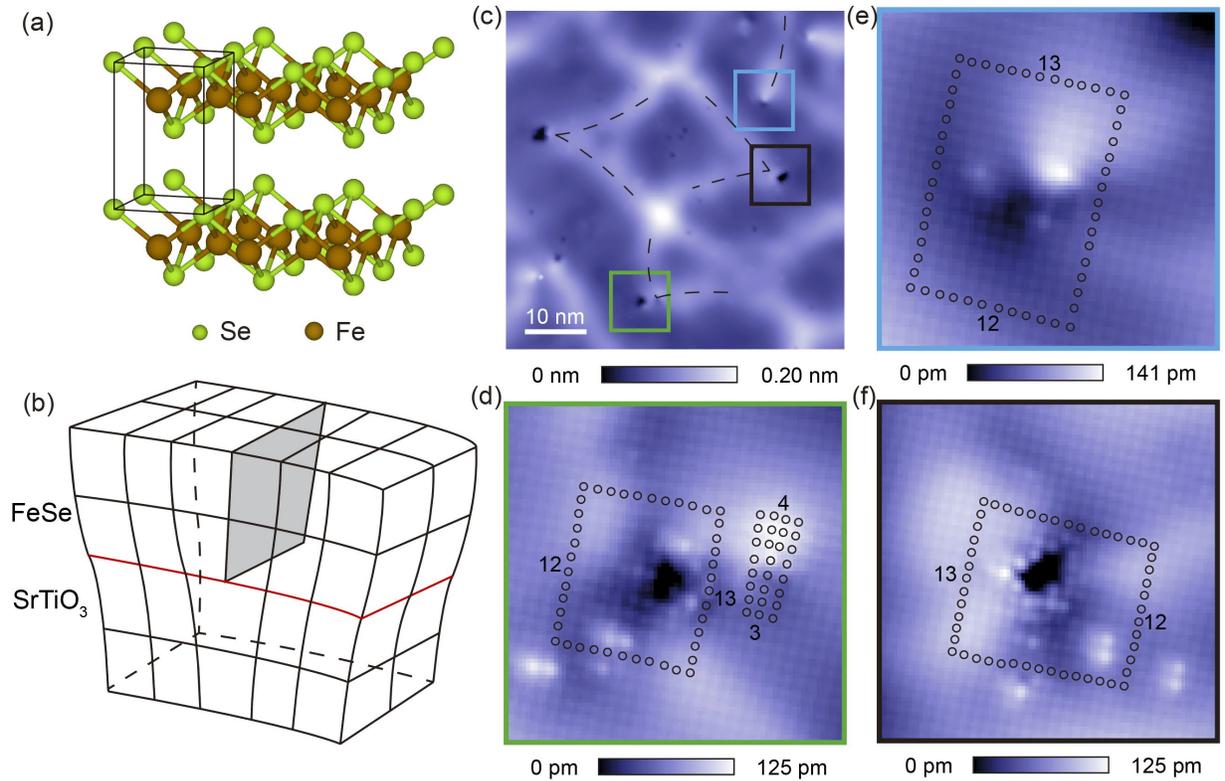

**Fig. 1** (a) Schematic of the crystal structure of 2 ML FeSe. Brown and green spheres denote Fe and Se atoms, respectively. (b) Schematic of an edge dislocation at the interface of FeSe/SrTiO$_3$. Shaded plane indicates the extra half-plane of atoms above the dislocation core. (c) STM topograph displaying a broken dislocation network observed at the surface of a 3 ML FeSe film. Dashed lines trace the dislocation lines that end at threading dislocations. (d)-(f) Magnification of the corresponding regions in c enclosing individual threading dislocations. STM setup condition: (c) $I_{set}$ = 10 pA, $V_{sample}$ = 1 V; (d)-(f) $I_{set}$ = 50 pA, $V_{sample}$ = -120 mV.

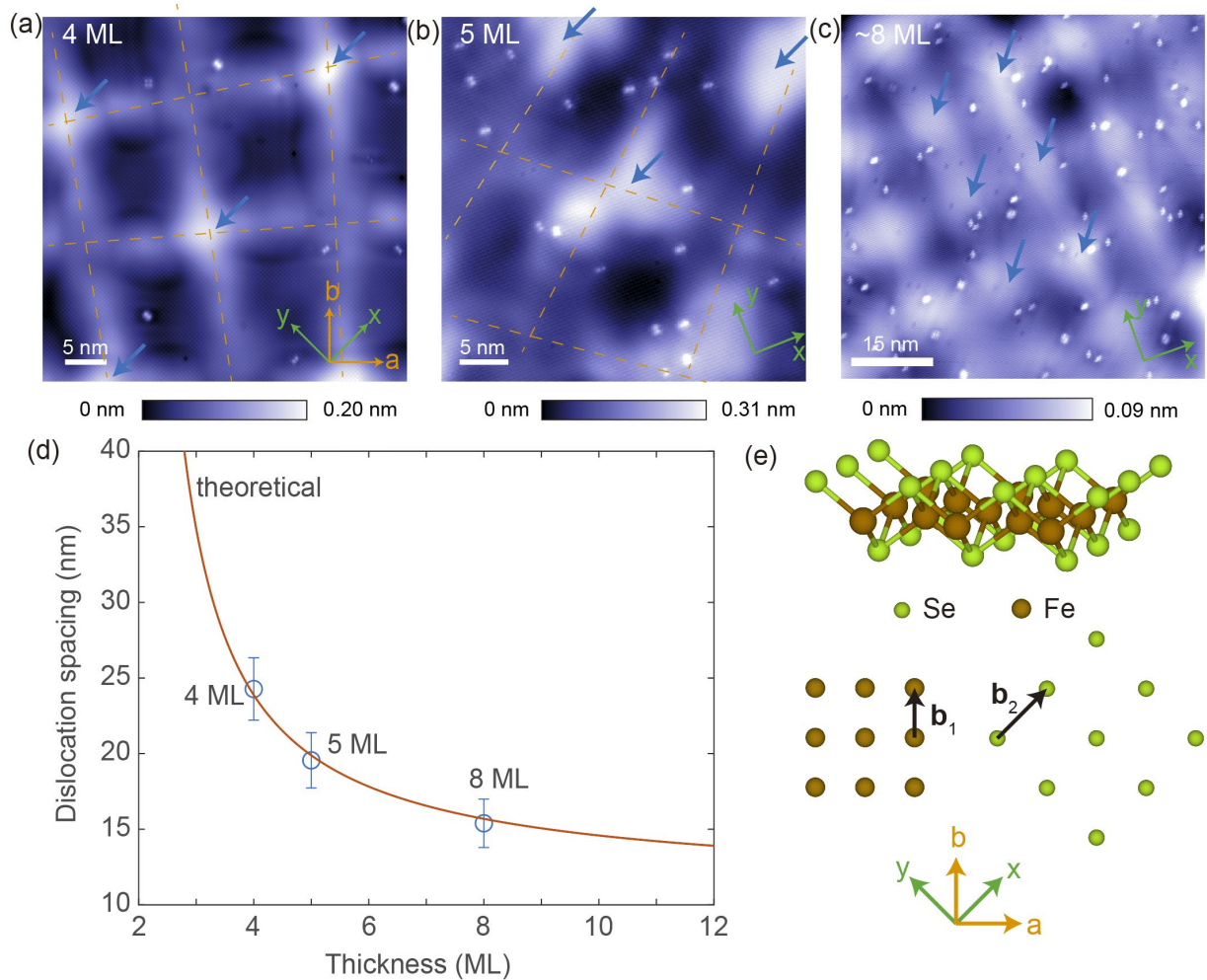

**Fig. 2** (a)-(c) STM topographs acquired at the surface of 4 ML, 5 ML and ~8 ML FeSe thin films, respectively, exhibiting an evolution of the dislocation network as a function of film thickness. While the network is oriented roughly along the a,b-axis (orange dashed lines) in the thinner film, it rotates in-plane by about 45 degrees towards the x,y-axis (blue arrows) in the thicker film. The average spacing between nearest-neighbor bright nodes (denoted by blue arrows) as a function of thickness is plotted in (d) (blue circles and error bars), which is fitted using the Matthews model (red curve). (e) Schematic of the two Burgers vectors $b_1$ and $b_2$. STM setup condition: (a) $I_{set}$ = 60 pA, $V_{sample}$ = 100 mV; (b) $I_{set}$ = 50 pA, $V_{sample}$ = 100 mV; (c) $I_{set}$ = 80 pA, $V_{sample}$ = 140 mV.

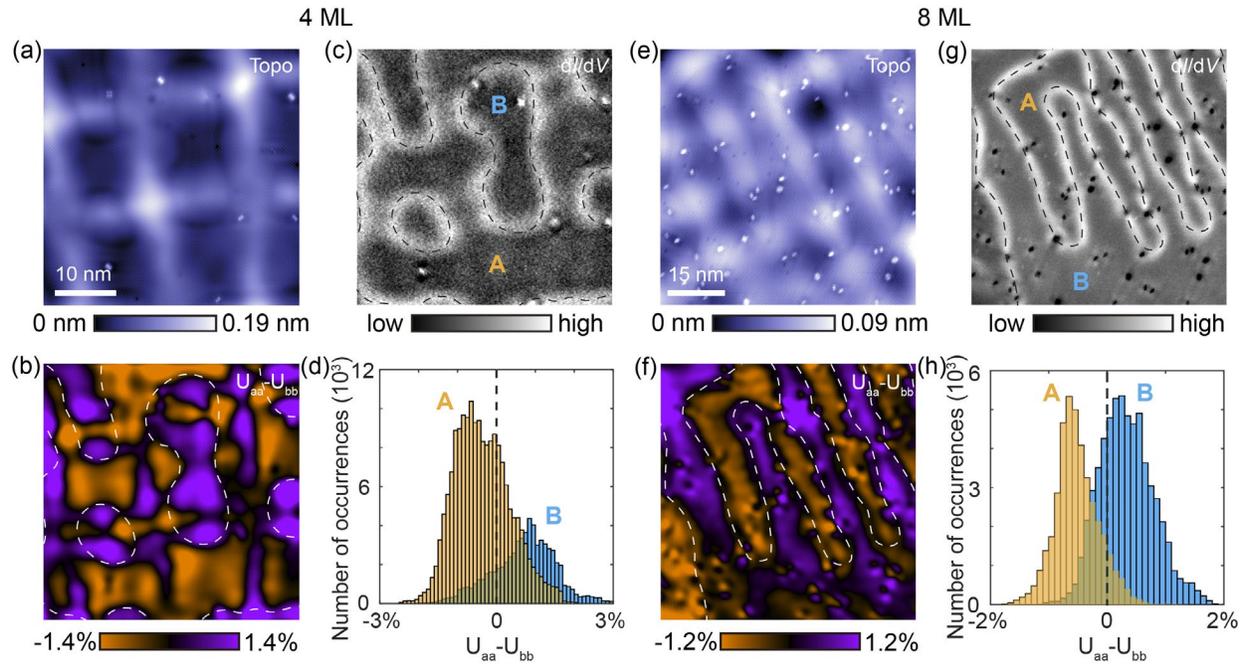

**Fig. 3** (a) STM topograph, (b) antisymmetric strain map $U_{aa}-U_{bb}$ (where $U_{ii} = du_i(r)/dr_i$, i=a,b and $u_i(r)$ is the displacement field) calculated from (a) (See Refs. [27,30,31,41] for more details), and (c) concomitant d$I$/d$V$ map. Dashed lines in (c) highlight the electronic domain boundaries and are superimposed on the antisymmetric strain map in (b). A histogram of $U_{aa}-U_{bb}$ within nematic domains A and B is shown in (d). (e)-(h) STM topograph, antisymmetric strain map, d$I$/d$V$ map and the strain histogram for the 8 ML film. STM setup condition: (a) $I_{set}$ = 60 pA, $V_{sample}$ = 100 mV; (c) $I_{set}$ = 110 pA, $V_{sample}$ = -100 mV, $V_{exc}$ = 5 mV; (e) $I_{set}$ = 80 pA, $V_{sample}$ = 140 mV; (g) $I_{set}$ = 80 pA, $V_{sample}$ = 140 mV, $V_{exc}$ = 10 mV.

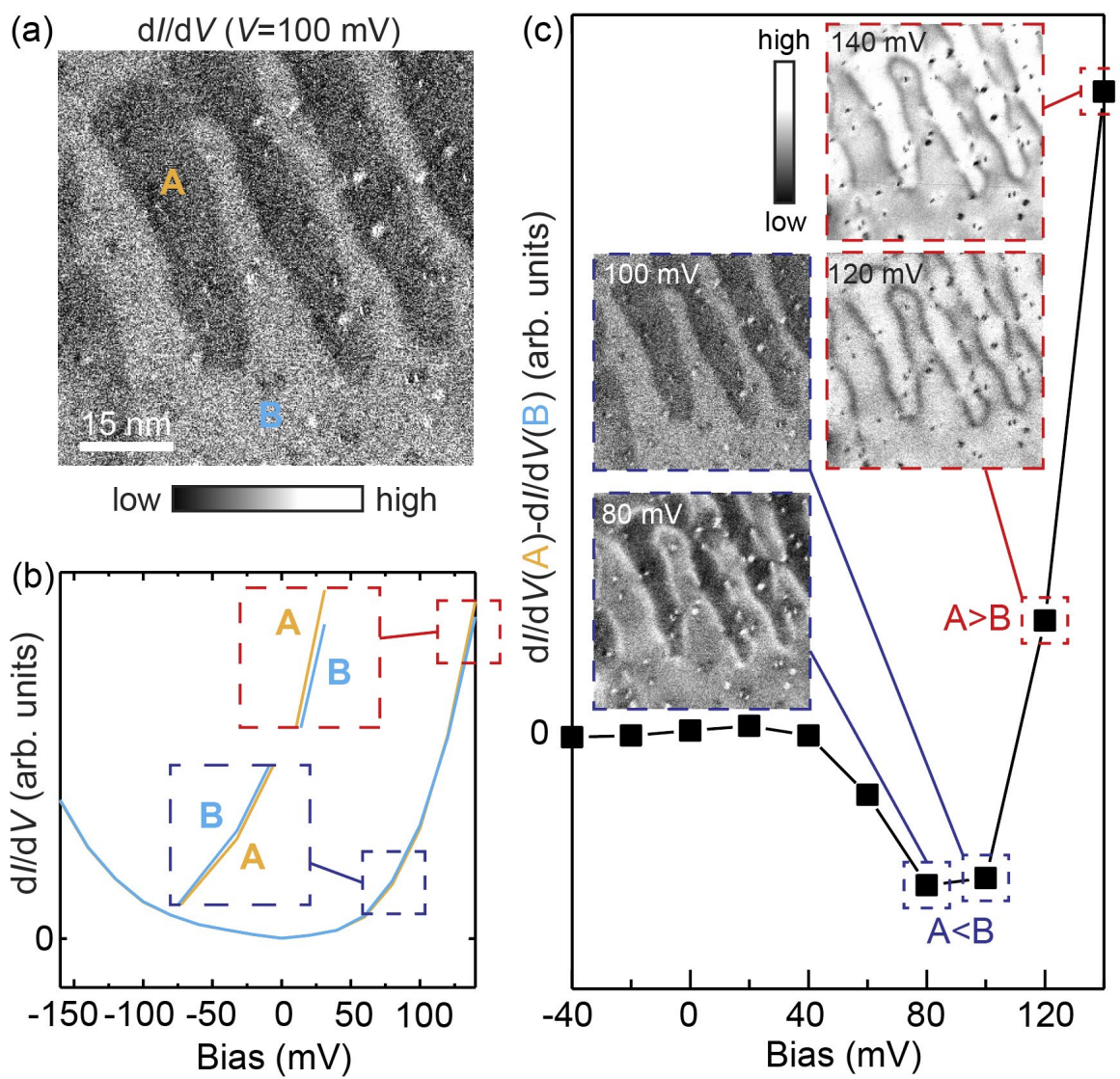

**Fig. 4** (a) d*I*/d*V* map acquired at the surface of an 8 ML film. A and B denote the two orthogonal nematic domains. (b) Average d*I*/d*V* spectra in domain A and B. Dark blue and red dashed boxes magnify the most pronounced differences between the two spectra. Their differential as a function of bias is displayed in (c), where relevant d*I*/d*V* maps with substantial nematic contrasts are shown as the insets. STM setup condition: $I_{set}$ = 80 pA, $V_{sample}$ = 140 mV, $V_{exc}$ = 10 mV.